\documentclass[iop]{emulateapj}
\usepackage{apjfonts} 

\usepackage{graphicx}
\usepackage{bm}

\makeindex
\citeindextrue

\received{December 19, 2013}
\accepted{April 2, 2014}
\journalinfo{Astrophysical~journal}
\submitted{}

\shorttitle{Recollimation Shock in BL Lacertae}
\shortauthors{Cohen et al.}

\defcitealias{Lis13}{L13}
\defcitealias{J05}{J05}

\begin{document}
\def\deg{\ifmmode^\circ\else$^\circ$\fi}
\def\n{\ldots}

\title{Studies of the Jet in BL Lacertae I. Recollimation Shock and
Moving Emission Features}

\author{M.~H. Cohen\altaffilmark{1}, 
D.~L. Meier\altaffilmark{2},
T.~G. Arshakian\altaffilmark{3,4},
D.~C. Homan\altaffilmark{5}, 
T. Hovatta\altaffilmark{1,6},
Y.~Y. Kovalev\altaffilmark{7,8}, M.L. Lister\altaffilmark{9},
A.~B. Pushkarev\altaffilmark{10,11,8}, J.L. Richards\altaffilmark{9},
\and
T. Savolainen\altaffilmark{8}}

\altaffiltext{1}{Department of Astronomy, California Institute of
Technology, Pasadena, CA 91125, USA; mhc@astro.caltech.edu}

\altaffiltext{2}{Jet Propulsion Laboratory, California Institute of 
Technology, Pasadena, CA 91109 USA} 

\altaffiltext{3}{I. Physikalisches Institut, Universit\"at zu K\"oln,
Z\"ulpicher Strasse 77, 50937 K\"oln, Germany}

\altaffiltext{4}{Byurakan Astrophysical Observatory, Byurakan  378433,
Armenia and Isaac Newton Institute of Chile, Armenian Branch}

\altaffiltext{5}{Department of Physics, Denison University, Granville,
OH 43023 USA}

\altaffiltext{6}{Aalto University Mets\"ahovi Radio Observatory,
Mets\"ahovintie 114, 02540 Kylm\"al\"a, Finland}

\altaffiltext{7}{Astro Space Center of Lebedev Physical Institute,
Profsoyuznaya 84/32, 117997 Moscow, Russia}

\altaffiltext{8}{Max-Planck-Institut f\"ur Radioastronomie, Auf Dem
H\"ugel 69, 53121 Bonn, Germany}

\altaffiltext{9}{Department of Physics and Astronomy, Purdue University,
525 Northwestern Avenue, West Lafayette, IN 47907, USA}

\altaffiltext{10}{Pulkovo Observatory, Pulkovskoe Chausee 65/1, 196140
St. Petersburg, Russia}

\altaffiltext{11}{Crimean Astrophysical Observatory, 98409 Nauchny,
Crimea, Ukraine}


\begin{abstract}
Parsec-scale VLBA images of BL Lac at 15 GHz show that the jet contains a
permanent quasi-stationary emission feature 0.26 mas (0.34 pc projected)
from the core, along with numerous moving features. In projection,
the tracks of the moving features cluster around an axis at position
angle -166.6\deg~ that connects the core with the standing feature.
The moving features appear to emanate from the standing feature in a
manner strikingly similar to the results of numerical 2-D relativistic
magneto-hydrodynamic (RMHD) simulations in which moving shocks are
generated at a recollimation shock.  Because of this, and the close
analogy to the jet feature HST-1 in M\,87, we identify the standing
feature in BL Lac as a recollimation shock.  We assume that the
magnetic field dominates the dynamics in the jet, and that the field is
predominantly toroidal.  From this we suggest that the moving features
are compressions established by slow and fast mode magneto-acoustic
MHD waves.  We illustrate the situation with a simple model in
which the slowest moving feature is a slow-mode wave, and the fastest
feature is a fast-mode wave. In the model the beam has Lorentz factor $\rm
\Gamma_{beam}^{gal}\approx 3.5$ in the frame of the host galaxy, and the
fast mode wave has Lorentz factor $\rm \Gamma_{Fwave}^{beam}\approx 1.6$
in the frame of the beam. This gives a maximum apparent speed for the
moving features, $\rm \beta_{app}=v_{app}/c= 10$. In this model
the Lorentz factor of the pattern in the galaxy frame is approximately
3 times larger than that of the beam itself.
\end{abstract}



\section{Introduction} 
\label{sec:intro}

BL Lacertae objects are a class of active galactic nucleus (AGN)
that contain a relativistic narrow outflow (a jet) aimed close to the
line-of-sight (LOS). This produces the characteristic features seen
at radio wavelengths: high brightness temperature, rapid variability,
and high polarization.  Many of the BL Lacs are gamma-ray emitters.
In high resolution radio images, some show a sharply bent or kinked jet
that changes on time scales as short as a year or less.  In some cases
bright features, or components, in the jet move downstream at nearly the
speed of light, $c$, giving them a proper motion $\mu$ with an apparent
speed $\beta_{app} > 1$ (in units of $c$) in the coordinate frame of
the galaxy. It is these rapid changes in the jet of the eponymous BL
Lacertae itself that we investigate in this series of papers.  In the
current paper we consider the kinematics of the components, and suggest
that the quasi-stationary component is a recollimation shock (RCS),
that the moving components emanate from the RCS, and that the moving
components are magneto-acoustic waves. In the next paper (Paper II,
in prep.) we show that the jet supports transverse waves as well, and
suggest that they are Alfv\'en waves. 

We need to distinguish between the pattern speed of a component and the
speed of the beam itself; i.e., the bulk speed of the plasma. It
is the latter that, through its Doppler factor, provides a relativistic
boost to the flux density, and it is the former that gives the apparent
motion. These speeds do not have to be the same, but it often is assumed
that they are (e.g. Lister et al. 2009). In \S\ref{move_comp_mhd} we
suggest that in BL Lac the moving components correspond to MHD waves
traveling longitudinally in a helical magnetic field, and thus that
the pattern and beam speeds are not the same.

We will need a value for $\theta$, the angle between the jet axis
and the LOS, and to choose one we have investigated values based on
statistics, and others based on observations of BL Lac as interpreted with
synchrotron-emitting models of the core.  The first statistical method
is based on the maximum observed apparent speed, $\beta_{app}=10.0$
(\citealt*{Lis13}, hereafter \citetalias{Lis13}), which gives an upper limit $\theta
< 11.4\deg$.  However, a source selected on the basis of its beamed
emission is highly unlikely to have $\theta$ near its extreme upper limit,
because the flux density is strongly deboosted there.  Roughly half of the
sources in a beamed flux density-limited sample will have $\rm\Gamma \sim
\beta_{app}$ and $\theta \sim 1/\Gamma$, while the most probable value
of $\theta$ is $\sim 1/2\Gamma$ \citep{LM97, Coh07}. Also, Pushkarev
et al. (2009, 2012) showed that the probability density function for
$\theta$ for a subset of MOJAVE sources that were detected with the
Fermi LAT has a peak near $2\deg$ and a median of $3\deg$. BL Lac is in
this group of gamma-ray emitters.  These studies suggest that $\theta$
is probably on the order of $3\deg$.

On the other hand, \citet{Hov09} have derived a value of $\theta=7.3\deg$
for BL Lac from the variability Doppler factor technique, and \cite{J05}
(hereafter \citetalias{J05}) have derived $\theta=7.7\deg \pm 1.9\deg$ using a
light-travel time argument. These are remarkably close. Thus measurements
and synchrotron theory give $\theta \sim 7.5\deg$, while probability
arguments suggest $\theta \sim 3\deg$.  Further observations that
affect $\theta$ are of the position angle (PA) of the jet, which for BL
Lac is variable and changes by up to $20\deg$ on time scales of 5 years
(\citetalias{J05}, \citetalias{Lis13}). When this is deprojected it means that $\theta$ itself
changes by a few degrees. \citet{CAM13} have studied these changes
with a precession model, and find that $\theta$ changes by about $\rm
4\deg~to~5\deg$, apparently without crossing the LOS.  In this paper,
when necessary, we will assume $\theta \approx 6\deg$, and that the
foreshortening is a factor of 10.

High-resolution observations at millimeter wavelengths show that,
on a scale of 0.2 mas, the PA of the inner jet of BL Lac is variable
\citepalias{J05}. These PA variations led to a suggestion by \citet[hereafter
S03]{Sti03} that the PA swings periodically, with a period of $2.3 \pm
0.3$ years. This time scale was verified by \citet{MD05} (hereafter MD05)
who also said that the period was not unique, and that a period of 13.1
years would fit as well. The \citet{CAM13} study yielded a precession
period of 12.1 yr.  In this paper we will show that on time scales of
1-12 years the PA is variable but not periodic.  The PA variations form
an important part of our story, as they appear to be connected to the
excitation of waves on the ridge line (Paper II).

Observational studies \citep{GPC00, LH05} have shown that the jets
of BL Lac sources can be highly  polarized, and that they have a bimodal
distribution of  electric vector position angle (EVPA) relative to the
jet axis. In most cases the EVPA is longitudinal; i.e. roughly parallel
to the axis, but in a small fraction of the cases it is transverse.
Longitudinal EVPA is often interpreted as arising from an optically thin
jet with a predominantly transverse magnetic field; although \citet{LPG05}
have emphasized that this is not necessarily the case. For a relativistic
beam the ray path in the beam frame is perpendicular to the axis when
$\theta=\theta_{crit}\equiv\sin^{-1}(1/\Gamma)$ and in this case, when
the field is helical with a high pitch angle and the jet is unresolved,
the EVPA is longitudinal.  However, the EVPA will vary as $\theta$
moves away from $\theta_{crit}$.  Transverse EVPA will be seen only
for small $\theta$ \citep{LPG05}.  If the jet is resolved, then for a
helical field there should be a transverse gradient of rotation measure
(RM) \citep{Bla93}.  \citet{GMC04} tested this by measuring the RM in four
BL Lacs, and they found indications of transverse gradients as expected
for a helical magnetic field. However, this result is controversial as
\citet{TZ10}, using stricter criteria, did not detect a gradient of RM
in four objects, including two of those studied by \citet{GMC04}. This
has been further studied by \citet{Hov12} with extensive simulations
of noise and other errors. They find a significant transverse gradient
in four sources, of which two, 3C 273 \citep{Asa02, ZT05} and 3C 454.3
\citep{Zam13}, have the signature expected for a helical field.

The evidence for a helical magnetic field in BL Lac comes from
the high polarization and from the longitudinal EVPA. \citet{OG09a}
have shown that its EVPA, corrected for rotation measure, remains
longitudinal around a bend, and that at 7.9 GHz the polarization rises
above $30\%$.  Since the maximum polarization that can be attained by
synchrotron radiation is 71\% in a uniform magnetic field, then in BL
Lac the field must be well-ordered, at least in those parts of the jet
where the polarization is high.  From these studies, and because we
can reasonably infer the existence of a helical field in 3C\,273 and
3C\,454.3, we assume that in the jet frame of BL Lac the magnetic field
basically has a helical shape, and that the pitch angle is not small.

The plan of this paper is as follows. In \S\ref{sec:obs} we discuss the
various bright features (components) in the jet (core, quasi-stationary,
moving) that are tracked by the MOJAVE program. In \S\ref{sec:cpt7}
we consider the quasi-stationary component near the core and why it
is likely that it is a recollimation shock.  In \ref{sec:other_AGN} we
extend this discussion to several other sources.  In \S\ref{move_comp_mhd}
we discuss the physical nature of the moving components in terms of
fast and slow magneto-acoustic waves.  \S\ref{sec:discussion} contains a
discussion of our results, and conclusions are in \S\ref{sec:conclusions}.

BL Lac is nearby for a blazar ($z=0.0686$) and our linear resolution is
high (1 mas corresponds to 1.29 pc in the galaxy). This provides a great
advantage to our study. The second great advantage we have is that BL
Lac varies rapidly, and phenomena including PA variation and wave motion
can be captured in a few years. 


\section{Observations}
\label{sec:obs}

The MOJAVE program \citep[Monitoring of Jets in Active Galactic Nuclei
with VLBA Experiments]{LH05} is conducting regular observations of about 300
sources with the VLBA at 15 GHz \citepalias{Lis13}. The observing cadence varies
from a few weeks for the most variable sources, to about 3 years for
slowly-changing ones. BL Lac varies rapidly and we are currently observing
it every one to two months. Here we present results from 55 epochs
between 1996.2 and 2013.0. For this paper we also have added results from
other programs whose results are in the VLBA archives, for a total of 110
epochs, all observed on the VLBA at a frequency near 15.3~GHz. See
\citetalias{Lis13}
for the most recent discussion of the MOJAVE program and its results,
and for references.

\begin{figure}[b]
\epsscale{1.0}
\plotone{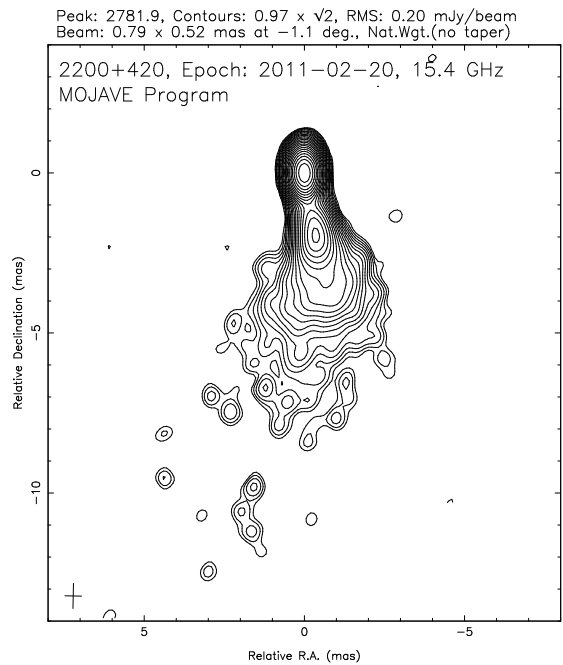}
\caption{Image of BL Lac at 15.4 GHz, epoch 2011.14.
\label{fig1}}
\end{figure}

We use multiple ``snapshot" images from the VLBA in this paper.
This involves observing BL Lac for a few minutes at each of a number of
widely-separated hour angles. The result has the full angular resolution
of the VLBA, but there still may be gaps in the uv-plane coverage, due
mainly to the wide separations of the fixed antennas.  Figure~\ref{fig1}
shows a 15 GHz image of BL Lac, made with the VLBA.  It has a dynamic
range of about 3000:1.  The jet lies at a small angle $\theta$ to the LOS
and the foreshortening is about a factor of 10.  The jet is actually long
and thin.  At the epoch of Figure~\ref{fig1} it is reasonably straight,
but at other epochs it is bent and sometimes kinked \citepalias{Lis13}.

\begin{deluxetable}{clcrrcrc}
\tablecolumns{8}
\tabletypesize{\scriptsize}
\tablewidth{0pt}
\tablecaption{\label{gaussiantable}Fitted Jet Components}
\tablehead{ \colhead {} &   \colhead {} &
 \colhead{I} & \colhead{r} &\colhead{P.A.} & \colhead{Maj.} &
\colhead{} &\colhead{Maj. P.A.}   \\
\colhead {I.D.} &  \colhead {Epoch} &
\colhead{(Jy)} & \colhead{(mas)} &\colhead{(\arcdeg)} & \colhead{(mas)} &
\colhead{Ratio} &\colhead{(\arcdeg)}  \\
\colhead{(1)} & \colhead{(2)} & \colhead{(3)} & \colhead{(4)} &
\colhead{(5)} & \colhead{(6)} & \colhead{(7)} & \colhead{(8)}}
\startdata
0 &1996-02-28 &    0.71&    0.00&    0.00&    0.07&    1.00&   \ldots \\ 
0 &1996-05-16 &    1.69&    0.00&    0.00&    0.00&    1.00&   \ldots \\ 
0 &1996-10-27 &    1.47&    0.00&    0.00&    0.04&    1.00&   \ldots \\ 
0 &1997-03-10 &    1.09&    0.00&    0.00&    0.00&    1.00&   \ldots \\ 
0 &1997-04-06 &    0.86&    0.00&    0.00&    0.00&    1.00&   \ldots \\ 
0 &1997-08-10 &    1.25&    0.00&    0.00&    0.06&    1.00&   \ldots \\ 
0 &1997-08-28 &    1.47&    0.00&    0.00&    0.04&    1.00&   \ldots \\ 
0 &1998-01-03 &    1.27&    0.00&    0.00&    0.03&    1.00&   \ldots \\ 
0 &1998-03-07 &    1.17&    0.00&    0.00&    0.07&    1.00&   \ldots \\ 
0 &1999-01-02 &    1.20&    0.00&    0.00&    0.06&    1.00&   \ldots \\ 
0 &1999-05-16 &    1.21&    0.00&    0.00&    0.10&    1.00&   \ldots \\ 
0 &1999-05-29 &    1.17&    0.00&    0.00&    0.09&    1.00&   \ldots \\ 
0 &1999-07-24 &    1.89&    0.00&    0.00&    0.05&    1.00&   \ldots \\ 
0 &1999-09-12 &    1.93&    0.00&    0.00&    0.23&    1.00&   \ldots \\ 
0 &1999-10-16 &    1.86&    0.00&    0.00&    0.00&    1.00&   \ldots \\ 
0 &1999-11-06 &    2.06&    0.00&    0.00&    0.00&    1.00&   \ldots \\ 
\enddata

\tablenotetext{b}{Fewer than 5 epochs, or $r>4$~mas}
\tablecomments{Columns are as follows:  (1) component identification number (zero indicates core component), (2) observation epoch, (3) flux density in Jy, (4) position offset from the core component in milliarcseconds, (5) position angle with respect to the core component in degrees,  (6) FWHM major axis of fitted Gaussian in milliarcseconds, (7) axial ratio of fitted Gaussian, (8) major axis position angle of fitted Gaussian in degrees.  }
\tablecomments{This table is available in its entirety in the online journal.}

\end{deluxetable}

We make a model of the source that consists of a set of Gaussian
components (circular when possible) and have made tests of the modelling
procedure, by comparing results obtained by people working independently.
In general, the stronger model components found with multiple snapshots
are unique.  Lister et al. (2013) showed that the accuracy of the
locations of the centroids of the components, important for this paper,
is about 1/5 the beamwidth, or about 0.1 mas EW by 0.2 mas NS, and for
isolated bright components it is half that. We generally adopt $\pm 0.1$
mas as the accuracy.

The strong source at the north end of the jet is the $core$, and a
prominent emission feature is seen at about 2 mas south of the core.
This feature is a blend of two of the components in the model for this
epoch. In Figure~\ref{fig1} the jet widens near $r=3$ mas and bends to
the SE. This is seen in every 15 GHz image of BL Lac.

If the observational epochs are sufficiently close together the model
components can be tracked without ambiguity, and they are generally seen
to move downstream.  We study each component and its motion carefully
to decide if it is robust.  This study includes an examination of the
statistics of independent $x$ and $y$ fits of linear and acceleration
terms to the motion.  Further weight is given to the physical reality of
a component when it has an identity that persists across a substantial
frequency range; e.g.  in \S\ref{sec:fast_cpt} we shall see that the
15--GHz component C2 has approximately the same location and motion as
the 43 GHz component S10 (S03).

Table~\ref{gaussiantable} gives the results of model-fitting
Gaussian components to the (u,v) data. This table is similar to Table 4
in \citetalias{Lis13}, with the data listed by component number rather than by epoch.
Most of the data here are the same as in \citetalias{Lis13}, but because the data
base is continuously being augmented, revisions are introduced into some
components. Thus the data here are not identical to the data in \citetalias{Lis13}. Our
final epoch for this paper is 23 December 2012.  The temporal coverage
prior to May 1999 is insufficient for robust cross-identification of
jet features, so for those epochs in Table~\ref{gaussiantable} we have
listed only the data for the core and for component 7, which are valid.

In this paper we restrict the discussion to physically plausible
components by selecting only those that satisfy the following criteria:
flux density $\rm{S}_{15}>20$ mJy and size (FWHM) $< 2.0$ mas, in addition
to noise and reliability requirements.  Furthermore, we flag and do
not use components in Table 1 that are seen at fewer than 4 epochs, or
have $r>4$~mas.  For July 2005 the core was constrained to be circular,
to obtain a useful fit for Component 7. This means that the component
positions in Table~\ref{gaussiantable} for July 2005 may not be compatible
with those for other epochs, in solving for the kinematic properties as
in Table~5 of \citetalias{Lis13}.

\begin{figure*}[t]
\includegraphics[width=\textwidth,trim=0.5cm 9cm 0.8cm 0cm]{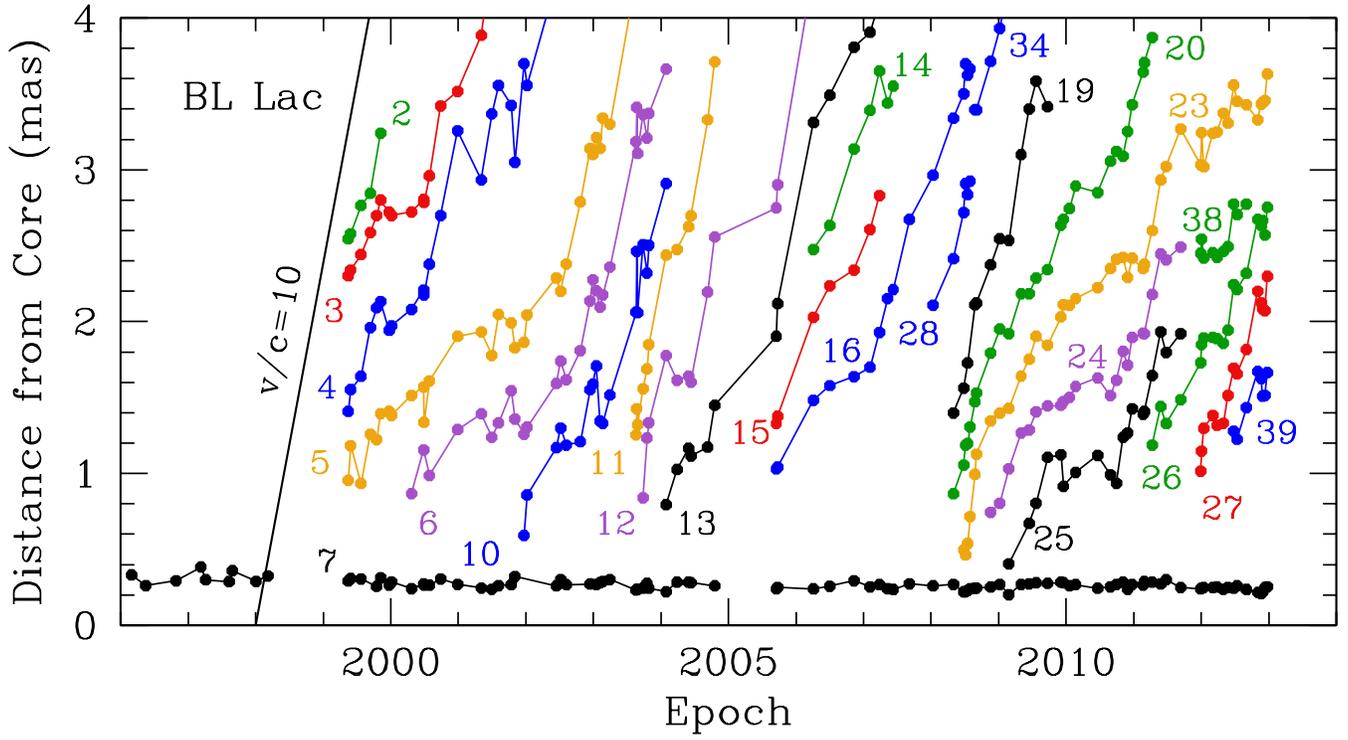}
\caption{Separation of the components from the core. The straight
black line has a slope corresponding to $\beta_{app}= 10$. 
\label{fig2}}
\end{figure*}

Figure~\ref{fig2} shows the radial motion of 24 components of BL Lac.
The straight line starting at $(1998,0)$ has a slope corresponding to
$\beta_{app}=10$, which is the value attained by component C1 although this
is not seen in Figure~\ref{fig2} because it lies at $r>4$ mas. The highest
value seen in Figure~\ref{fig2} is $\beta_{app}=9.2$, for component C11
\citepalias{Lis13}.

\subsection{The Core}
\label{sec:core}

The core of a jetted radio source is generally defined as the compact
flat-spectrum component seen at one end of the jet; it appears at
nearly every epoch and often is assumed to be stationary on the sky.
In Figure~\ref{fig1} the core is at the north end of the image, and is
the reference point for all other components.  However, the absolute
position of the core can depend on the observing frequency, since
usually it is regarded as the photosphere; i.e., the $\tau \approx 1$
region in a smoothly-varying jet (see e.g. Sokolovsky et al. 2011;
O'Sullivan \& Gabuzda 2009b, Kovalev et al. 2008). The measured core
shift in BL Lac, between 8 and 15 GHz, is of order $\rm 50~\mu as$
\citep{Pus12}. Further, the core may prove to be a compound structure
when viewed at a shorter wavelength, with more resolution and less
opacity. This is the case for BL Lac, as we now show.

Independent 43 GHz VLBA observations of BL Lac were made by \citetalias{J05} and by
MD05 over partly-overlapping intervals between 1998 and 2003.
\citet{J05} show
that BL Lac had three permanent components in the inner region: the
43 GHz core itself and two components to the south, labeled A1 and A2
(see \citetalias{J05}, Figure~16). \citet{MD05} obtain a similar result; they see the
same three components and call them C1, C2, and C3. S03 call the two
northern components C1 and C2.  To avoid confusion we will exclusively
use the designations A0, A1, and A2 for the 43-GHz components, to refer
to the core and the two components labeled A1 and A2 by \citetalias{J05}. At 43 GHz
A1 is substantially stronger than A0. We assume that A0 is the 43 GHz
core of BL Lac, and A1 and A2 are stationary shocks in the jet. It is
unlikely that A1 is the core and A0 is part of a counter-jet, because the
Lorentz factor of the beam is of order 3 or more, and therefore
there is an expected ratio of $10^3$ or more between the flux densities
of the jet and the counter-jet.

\citet{J05} list the separations of A1 and A2 from A0 as 0.10 and 0.29
mas, respectively (\citetalias{J05}, Table~5).  S03 show that the PA from A0 to A1 is
variable, between about $-153\deg$ and $-190\deg$.  It was this swing
in PA that led them to suggest a periodicity.  At 15 GHz the angular
resolution of the VLBA is insufficient to separate A0 and A1 and together
they form the core at 15 GHz.  The variable PA between A0 and A1, however,
does mean that the 15 GHz core moves on the sky. The motion depends
on the 15 GHz flux density ratio between A0 and A1; this is unknown
but at 43 GHz A1 is substantially stronger than A0 (\citetalias{J05}, MD05). In
any event, the position shift will mainly be in the RA direction; the
sky motion will be $\rm\le \rm 60~\mu as~y^{-1}$.  This is consistent
with the proper motion measured at 8 GHz during 1994--1998: $\rm 20.4
\pm 6.6~\mu as~y^{-1}$, at $\rm{ PA}=-16 \pm 16^\circ$ \citep{Moo11}.
The accuracy of our position measurements is estimated as $\pm 0.1$
mas so we ignore the possible shifts in the apparent core position.

\subsection{The Quasi-Stationary Component}
\label{sec:slow_cpt}

\begin{figure}[t]
\includegraphics[width=0.5\textwidth,trim=0.3cm 0cm 4cm 0.7cm]{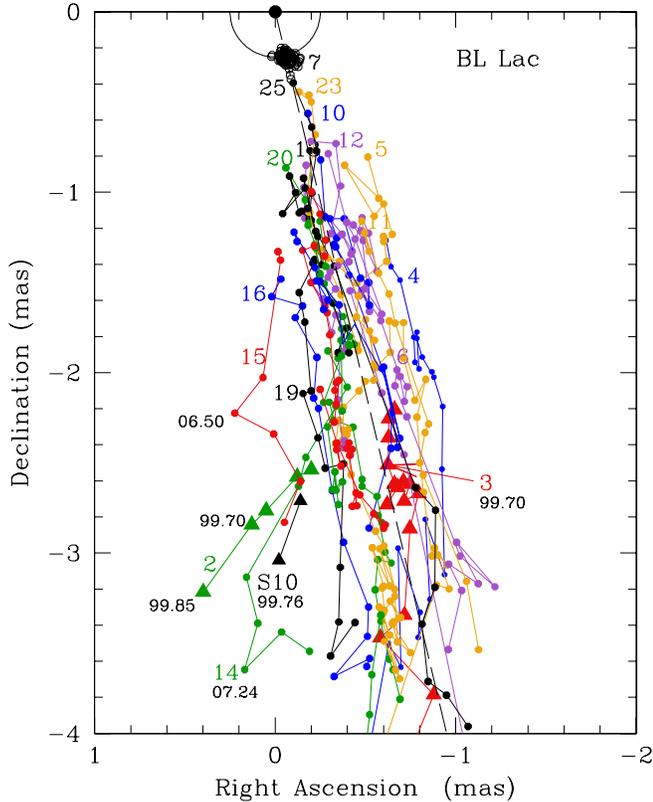}
\caption{RA-Dec tracks for the components, relative to the core at
(0,0). The circle at $r=0.25$ mas is for convenience.  The cluster of
black points is component 7, which we identify as a recollimation shock.
The E-W uncertainty is $\pm 0.1$ mas.  The dashed line at $\rm PA =
-166.6\deg$ is an axis that connects the core and the mean of component
7. The solid black triangles are for the 43 GHz component S10. See text.
\label{fig3}}
\end{figure}

At 15 GHz BL Lac contains a quasi-stationary component, C7, seen in
Figure~\ref{fig2} at about 0.26 mas from the core. In Figure~\ref{fig3},
an RA-Dec plot, C7 is seen as the tight cluster of points with a centroid
at $r = 0.26$ mas and position angle $\rm{PA} =-166.6^\circ$. This is close
to the location of the 43 GHz component A2 from its core, $r=0.29$ mas,
$\rm PA \approx -166\deg$.  We therefore identify C7 with the 43 GHz
component A2. The small difference in radius at the two frequencies can
be attributed to the compound nature of the core as described above.

The 15 GHz restoring beam of the VLBA for BL Lac, calculated with
naturally-weighted data, is roughly $0.9\times 0.6$~ mas (FWHM)
at PA~$=-9\deg$, so that the separation between the core and C7 is
less than the beam size. Such super-resolution generally leads to a
non-unique model, whose utility depends on the SNR and on the true level
of complication in the field.  In our case the SNR is high, and from
the 43-GHz observations we know that there is only a small number of
compact components in the field.  Hence we have an $a~priori$ case that
superresolution is reasonable.  We also justify our superresolution by the
good agreement for the positions of components C7 and A2 at 15 and 43 GHz.

\begin{figure}
\includegraphics[width=0.5\textwidth,trim=0cm 9.5cm 0cm 0cm]{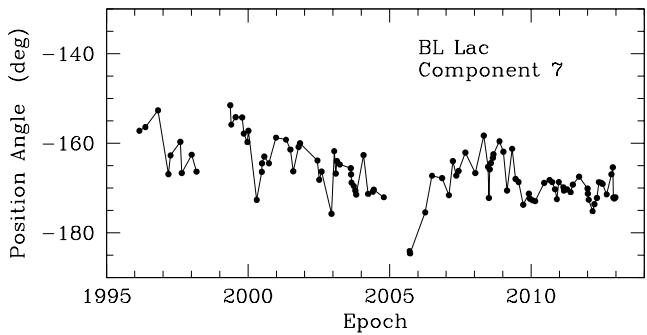}
\caption{Position Angle $vs$ Epoch for Component 7.
\label{fig4}}
\end{figure}

In Figure~\ref{fig4} we show the PA of C7 relative to the core as a
function of time.  The PA is stable after about 2009.5, and the scatter
there is an indication of the noise in the measurements, which appears
to be about $\pm 3\deg$. There is no periodicity that is apparent,
on scales from 1 to 12 years, but longer periods are not excluded.
MD05 (their Figure 3) show the combined S03 and MD05 data at 43 GHz
for the PA close to the core. Their data closely match ours for the
overlap period, 1999.0--2002.0, with a lag of a few months for the 15 GHz
data. This good agreement is further justification for superresolution,
and for using $\pm 0.1$ mas as a conservative estimate of the error in
the positions of the components.

It is not unusual to see a quasi-stationary component in a jetted
source, but the tightness of the cluster of points for C7 is remarkable.
In Figure~\ref{fig4} the RMS scatter in the RA and Dec positions, relative
to the mean, is $\rm 30~\mu as~and~28~\mu as$, respectively. The net vector
proper motion is $\rm 3.8\pm 0.6~\mu as~y^{-1}$ at $\rm PA=82\deg \pm
10\deg$ \citepalias{Lis13}.

\subsection{The Moving Components}
\label{sec:fast_cpt}
The slope $\mu$ for a component in Figure~\ref{fig2} gives its apparent
radial speed, calculated as $\beta_{app}=\mu RD(1+z)$ where
$ \beta_{app}$ is relative to $c$ the speed of light, $\mu$ is in
$\rm mas~y^{-1}$, D is the angular diameter distance in light-years,
and R converts mas to radians.  See \citet{Hom09}, who discuss the
definition of speed and acceleration.  It is clear that in BL Lac there is
a wide distribution of speeds, but it is the fastest one that is usually
quoted. This honor goes to component C1 which has the value $\beta_{app}
= 10.0$ \citepalias{Lis13}, although this component has $r > 4$ mas and is outside 
the purview of this paper. 

Nineteen of the moving 15 GHz components are shown in
Figure~\ref{fig3}. Five of the components in Figure~\ref{fig2}
are not shown because they have only 4 or 5 epochs each and merely add
to the general group of unmarked sources near the axis. The dashed
line at $\rm{PA}=-166.6\deg$ runs from the core through the mean of C7.
This axis is roughly parallel to the tracks of the moving components,
which surround it in a somewhat symmetrical fashion, except for components
C2, C14, and C15.  The moving components all appear to start downstream
of C7.  There is no indication that there are moving components upstream
from C7, although we have little angular resolution there.  This bears
a close resemblance to the moving shocks generated by the stationary
recollimation shock in the numerical simulations by \citet{Lin89}, which
we discuss in \S~\ref{sec:theory}.  At 43 GHz the situation is similar,
except for one episode in 2005 when a moving component was seen traveling
superluminally from the core to C7 \citep{Mar08}. This suggests that the
43 GHz core itself may possibly be an RCS, and that yet-higher frequency
studies may be needed to sort out this situation.

Most of the moving components start near $r=0.8$ mas, with only
three, C10, C23, and C25 clearly appearing before $r=0.6$ mas. This may
be due to sensitivity, with most components being born weak and only
rising to a measureable flux level after traveling several pc. Or,
it might represent a jet phenomenon. This shortage of flux close to
component 7 has been discussed by \citet{Bac06}.  Note also that
components C10, C23, and C25, which are close to the core, start close
to the axis at $\rm PA=-166.6\deg$, whereas those that start 
farther downstream are spread across 0.5 mas surrounding the axis.

\citet{Tat09}, using VLBA archival data at 8 and 15 GHz,
described the motions in BL Lac as a succession of ballistic components
traveling through a narrow stationary window to the south, running
from about 1 to 2 mas. That description is generally consistent with our
Figure~\ref{fig3}.  \citet{CAM13} have published a study of BL Lac
that includes an RA-Dec plot like that in Figure~\ref{fig3} but for 311
points at 6 frequencies from 5 to 43 GHz, all taken from the literature,
and all obtained with VLBI between 1979 and 2008.  Their Figure 1 is
similar to our Figure~\ref{fig3}, but their individual components are
not followed in time.

On the west side of the tracks in Figure~\ref{fig3}, components C4 and C5
form an envelope that is roughly parallel to the axis. On the east side,
however, there are several components, C2, C14 and C15,
with large transverse excursions.  The 43-GHz component S10 similarly
has a large eastern excursion, and two points from the data listed by
S03 are included in Figure~\ref{fig3}, as the solid black triangles.
The final points for C2 and S10 are labeled with their epochs. They
are close enough together in both space and time that we identify them
as the same component, with the spatial difference probably due to a
spectral gradient. The independent 15 and 43 GHz observations show that
C2 and S10 are real and not artifacts of the observations or reductions.
Component 14 occupies the SE region taken by C2, but 7 years later.

Component C2 is contemporaneous with the early part of component C3,
which is labeled with the arrow in Figure~\ref{fig3}, and shown as the
large red triangles.  C3 stays close to the axis, while C2 moves to the
SE. Unfortunately, we have only limited 15 GHz observations immediately
prior to 1999, so we cannot see if C3, moving radially, spawned C2 that
moved with a large transverse component.

These excursions of the moving components away from the axis are
all on the east side, which is in the direction of the general bend in
the source seen at $r>3$ mas in Figure~\ref{fig1}.  The character
of the source changes near 3 mas: the narrow jet becomes a broad low-surface 
brightness extended region without a sharp ridge. The components
there may not be on a pressure maximum, and may have a
different character from the ones on the narrow ridge closer to the core.

\begin{deluxetable*}{lllcccccc}
\tablecolumns{9}
\tabletypesize{\scriptsize}
\tablewidth{0pt}
\tablecaption{\label{distancetable}Distance to Recollimation Shock}
\tablehead{\colhead {Name} &   \colhead {z} &
\colhead{Class}&\colhead{pc/mas}&\colhead{theta}&\colhead{Dist to Shock}&\colhead{$\rm\log M_{BH}$}&\colhead{$\rm\log R$}&\colhead{Ref}}
\startdata
  BL~Lac      & 0.0686  & BLL   & 1.29  &  6 & 0.26 & 8.2  & 5.6 &  1,2 \\
  M~87        & 0.00436 & FR~I  & 0.08  & 13 & 860  & 9.5  & 6.0 &  1,3,4 \\
  3C~120 S1   & 0.033   & FR~I  & 0.65  & 16 & 0.7  & 7.8  & 5.7 &  5,6 \\
  3C~120 C80  &  \n     & \n    & \n    & \n & 80   & \n   & 7.8 &  6,7 \\
  3C~273      & 0.158   & FSRQ  & 2.70  &  6 & 0.15 & 9.8  & 4.1 &  8,9 \\
  3C~390.3 S1 & 0.0561  & FR~II & 1.09  & 50 & 0.28 & 8.6  & 4.3 &  10,11 \\
\enddata

\tablecomments{Columns are as follows:  (1) common name, (2) redshift, (3) BLL $=$ BL Lacertae object, FR~I $=$ Fanaroff-Riley type I,
FSRQ $=$ flat-spectrum radio quasar, FR~II $=$ Fanaroff-Riley type II, (4) linear scale, (5) angle of inclination of jet to the LOS, (6) projected distance to shock in mas, (7) log mass of the black hole, (8) log deprojected distance to shock in gravitational radii, (9) references}

\tablerefs{
(1) this paper,
(2) \citet{WU02},
(3) \citet{Che07},
(4) \citet{Geb11},
(5) \citet{Leo10},
(6) \citet{Gri12},
(7) \citet{Agu12},
(8) \citet{J05},
(9) \citet{PT05},
(10) \citet{WPM99},
(11) \citet{Ars10}
}

\end{deluxetable*}

\section{The Quasi-Stationary Component as a Recollimation Shock}
\label{sec:cpt7}

A ``master" recollimation, or reconfinement, shock (RCS) is a natural,
quasi-stationary feature of an MHD wind/jet that is expected to form
beyond the ``final critical point" -- the place where the accelerating
and collimating flow becomes causally disconnected from the central
engine. Sometimes called the modified fast point (MFP), the final critical
point was ignored in early studies of jet acceleration \citep{BP82,
LCB92}; and even the more recent \citet{VK03}, who assumed that the
MFP lay infinitely far from the central black hole.  However, the MFP
has been studied extensively by \citet{Vla00} and by Polko et al. (2010,
2013a, 2013b) who showed that it can lie at a finite distance within a
galactic nucleus (e.g., $\lesssim 10^6\,r_g$ or parsec scale, where $r_g =
\rm{G M_{BH}}/ c^2$ is the black hole gravitational radius).  At the MFP
the jet has a strong, toroidally-dominated magnetic field, the forward
flow is highly {\em super-magnetosonic} and converging, and even that
flow {\em toward} the central jet axis exceeds the magnetosound speed.
Within less than a magnetosound crossing time the flow will converge
beyond the final critical point into a strong compressive recollimation
wave or shock.

\subsection{Theoretical MHD Simulations of the RCS}
\label{sec:theory}

Helical field dominated, super-magnetosonic flow like that expected
beyond the MFP has been studied extensively with both non-relativistic
and relativistic, two-dimensional MHD simulations \citep{CNB86, Lin89,
Kom99, KC01}.  A strong MHD pinch shock indeed forms quickly, and just
beyond it a magnetic chamber that alternately opens and closes, ejecting
new jet components forward at trans- (not super-) magnetosonic speeds.
Since, to our knowledge, super-magnetosonic, helically-dominated MHD
jets have not yet been studied in three dimensional simulations, the
details of these results must be considered tentative only.  However,
it is clear from all the simulations that the RCS dramatically changes
the character of the flow, giving birth to a new and more stable {\em
trans}-magnetosonic jet from the previously unstable super-magnetosonic
pre-shock jet.  See \citet[\S15.2.2.2]{Mei12}, \citet{Mei13} for a more
detailed discussion.

\subsection{Comparison with M\,87}
\label{sec:comparison}

The jet of M\,87 contains a quasi-stationary emission region called
HST-1, $\ge 120$ pc from the BH, and several superluminal components
$(\rm \beta_{app,max}=4.3)$ downstream from HST-1 \citep{Che07, Cha10}.
Cheung et al., also \citet{Sta06} and \citet{BL09} discuss HST-1 as
a reconfinement or recollimation shock generated by a change in the
gradient of the external pressure, and they suggest that the superluminal
components are generated in the HST-1 shock.

Component 7 of BL Lac is analogous to HST-1 in M\,87. It is
quasi-stationary, and, as shown in Figure~\ref{fig3}, the superluminal
components appear to emanate from it, although there is not enough angular
resolution to state definitively that none of the moving components
starts upstream of C7. At 43 GHz there is more angular resolution. \cite{J05}
(their Figure~16) show three superluminal components that all start
near component A2 (same as the 15 GHz component C7).  Additionally,
\citet{Mar08} report an exceptional event in 2005 in which a 43-GHz
component appeared to start slightly upstream of the core, and was
tracked, moving superluminally, to component A2.

Table~\ref{distancetable} summarizes the details of the shocks in these
sources, and the others in \S\ref{sec:other_AGN}.  The objective of the
Table is to compare the distances (in gravitational radii) of the putative
RCS from the SMBH, in column 8. These values have a large uncertainty,
perhaps 0.6 dex.

To calculate the distance we must assume values for the inclination
angle $\theta$ and for $\rm M_{BH}$.  For BL Lac we assume $\theta
\approx 6^\circ$ as described above, and we take 
$\rm\log M_{BH}/M_\sun =8.2$
from \citet{WU02}.  This is lower than the typical values of 
$\rm M_{BH}/M_\sun$
for BL Lacs reported by \citet{FKT02}, 
$\rm log M_{BH}/M_\sun\sim 8.6$, but
their sample was small and had a large variance, and did not include
BL Lac.  Using $\rm\log M_{BH}/M_\sun =8.2$ 
gives $\rm{R} \sim 4\times 10^5\,
r_g$. This value is uncertain enough that there is no need to consider
adding the distance of the core from the BH itself.

For M\,87 \citet{Che07} derive a limit to the deprojected distance
of HST-1 by using the maximum observed superluminal speed and the
corresponding maximum angle to the LOS, $26\deg$. In our opinion
the maximum angle is unrealistic, and it is preferable to estimate
the distance by using the angle that minimizes the Lorentz factor,
$\theta=13\deg$. M\,87 is not selected on the basis of its beamed flux
density as is BL Lac, and it would be inappropriate to use a further
reduction in the angle. Thus we adopt R=300 pc as the de-projected
distance from the BH to HST-1.  The mass of the BH in M\,87 is estimated
as $\rm 6.6 \times 10^9 \, M_\sun $ \citep{Geb11}, giving a distance to
HST-1 of about $10^6 \, r_g$.  Component 7 in BL Lac and HST-1 in M\,87
are both quasi-stationary, and they are at similar gravitational distances
from their black holes. In both objects the major superluminal motions
are seen downstream of the stationary component.  HST-1 is presumed to
be a recollimation shock, and we suggest that component 7 is one, also.

In the \citet{Lin89} strong-field, super-magnetosonic simulations,
successive pairs of shocks appear downstream of the recollimation
shock.  They move in a narrow ``nose cone" (a post-RCS, trans-magnetosonic
flow) where the pressure is much higher than it is in the surrounding
cocoon, due to plasma confinement by a toroidal magnetic field.  We
have suggested that C7 in BL Lac is an RCS, and we now suggest that
the fast components downstream of C7 are {\em moving} shocks in
this post-RCS flow.

The comparison of the BL Lac observations with the \citet{Lin89}
simulations is striking, and lends credence to the idea that C7, like
HST-1 in M\,87, is indeed a recollimation shock.  However, some details,
including the transverse excursions seen in Figure~\ref{fig3}, have no
analog in the axisymmetric simulations.

\subsection{Quasi-Stationary Components in Other AGN}
\label{sec:other_AGN}

Many AGN show quasi-stationary components in parsec-scale images; see
e.g.\ \citetalias{J05}, Figure~16 and \citetalias{Lis13}, Figure~3. In this section we briefly discuss such
components in two radio galaxies, 3C\,120 and 3C\,390.3, and one quasar,
3C\,273. Their details are in Table~\ref{distancetable}.  3C\,120, like
M\,87, is an FR\,I radio galaxy and these objects, according to standard
unification theory, are the parent population of the BL Lacs.  MOJAVE
images at 15 GHz show that 3C\,120 has a quasi-stationary component,
called S1 by \citet{Leo10}, 0.7 mas from the core.  Taking $\theta\approx
16\deg$ \citep{Agu12} and $\rm M_{BH} \approx 6.7\times 10^7\,M_\sun$
\citep{Gri12} gives $\rm{R} \approx 5\times 10^5\, r_g$.

In 3C\,120 numerous radio components seen at 15 GHz appear to start
at or near S1 and move superluminally downstream \citep{Leo10}. At 43
GHz, however, \citep{Gom01, J05} it is seen that several components
start close to the core and pass through S1. This difference may be due
to the differing angular resolutions, but possibly also to the spectra
of the components, which may be strongly inverted when the components
first come out of the core.  In addition, 3C\,120 had 5 or 6 optical
flares that were closely correlated in time to the back-extrapolated
passage of the 15 GHz superluminal components through the component
S1 \citep{Leo10}.  The flares were not well-correlated with the times
when the back-extrapolated superluminal components were at the core.
During the appropriate time period there were 7 radio components, so it
appears that the correspondence between the optical flares and the times
when the superluminal radio components were at S1 is good. Le\'on-Tavares
et al. discuss the statistics of this association.

Table~\ref{distancetable} shows that S1 in 3C\,120 is at a distance
from the BH that is comparable to those of HST-1 for M\,87 and C7 for
BL Lac; superluminal components from the nearby core move through it,
and it appears to have a connection with optical flare events. We agree
with \citet{Leo10} that S1 is probably an RCS.

3C\,120 also has a quasi-stationary component 80 mas from the core,
and \citet{Agu12} suggest that it is an RCS. However, this component
is 1-2 orders of magnitude farther from the BH than are the other RCS
in Table~\ref{distancetable}, and furthermore the main superluminal
components in 3C\,120 lie far $upstream$ from C80.  We therefore suggest
that component C80 in 3C\,120 is of a different character from HST-1 in
M\,87, C7 in BL Lac, and S1 in 3C\,120.

3C\,273 is a flat-spectrum radio quasar that has a quasi-stationary
component in its jet at 0.15 mas from the core, seen in VLBA images
at 43 GHz \citep{J05, Sav06}.  Estimates of $\theta$ in the literature
range from 3.3\deg~\citep{Hov09} to 10\deg~\citep{Sav06}; for this paper
we will use $\theta \approx 6\deg$ \citepalias{J05}. Using $\rm M_{BH} \approx
6.6\times 10^9\,M_\sun$ \citep{PT05} gives $\rm{R} \approx 1.2\times
10^4\,r_g$. This value for $\rm{R}/r_g$ is about two orders of magnitude
smaller than the values for BL Lac, M\,87 and 3C\,120. The low distance
estimate does not preclude its being an RCS, but it may be of a different
character from those discussed above.  Another difference is that, like
many radio loud quasars, 3C\,273 has a predominantly transverse EVPA in
the jet, whereas the BL Lac objects have a predominantly longitudinal
EVPA.

3C\,390.3 is an FR\,II radio galaxy with a jet that has
$\Gamma \sim 2$ and $\theta \sim 50\deg$ \citep{Ars10}. \citet{Ars10}
combined their 15 GHz VLBA observations with observations made by the
MOJAVE group, for a total of 21 epochs during 1994-2008. They identified
a quasi-stationary component (S1) 0.28 mas from the core.  Taking $\rm
M_{BH} \sim 4\times 10^8\, M_\sun$ \citep{WPM99} gives $\rm{R}/r_g \sim
2\times 10^4$. This value is comparable to the value for 3C\,273, and
smaller than the others.  Again, this difference does not disqualify S1
from being an RCS, but it may have a different character from those in
M\,87, BL Lac, and 3C\,120.

In 3C\,390.3 numerous 15 GHz components appear to arise within a few
tenths of a parsec of S1, and move downstream with $\rm \beta_{app}
\sim 1$.  During 1994-2008 3C\,390.3 had 8 optical or UV flares, and
their peaks occurred close to the times when the back-extrapolated moving
components were at the location of S1 \citep{Ars10}.  The flares were
not well-correlated with the times when the back-extrapolated moving
components were at the core. This behavior is like that in 3C\,120.
The flares appear to be connected with the passage of the superluminal
components through S1.

\section{The Moving Components as Propagating Magneto-Acoustic MHD Waves
or Shocks}
\label{move_comp_mhd}

In \S\ref{sec:intro} we briefly reviewed the subject of
polarization in BL Lacs to justify our assumption that the magnetic
field in the jet is in the form of a helix with a high pitch angle. We
also assume that the magnetic pressure is stronger than the plasma
fluid pressure.  This allows us to invoke MHD to explain the observed
phenomena. The conditions in the post-shock jet in the MHD
recollimation shock simulations discussed in \S\ref{sec:theory} are
surprisingly similar to the actual state of the BL Lac jet downstream
from Component 7.  In the simulations, the RCS is essentially
non-dissipative, the new trans-magnetosonic jet still has a strong
magnetic field (strong in the sense that the field contributes
significantly to the jet dynamics), and the field retains its helical
configuration.  Furthermore, such a jet should display both kinds of
classical magneto-acoustic waves (and shocks) --- the MHD slow and fast
modes -- particularly along the jet axis.  These waves are disturbances
in the longitudinal direction, or P (pressure) waves.  In this section
we discuss slow and fast magneto-acoustic waves that, we suggest, form
the moving components in the jet.

\subsection{Slow MHD Waves and Shocks in the BL Lac Jet}

Classical slow MHD waves are the lower branch of the magneto-acoustic
modes.  They compress only the jet plasma and not the magnetic field,
and they propagate with a phase speed of only 
\begin{eqnarray} \label{VS}
V_S & = & \pm \, \left\{ \frac{1}{2} \left[ c_{ms}^2 - \left(c_{ms}^4 -
4 \, c_s^2 \, V_A^2 \, \cos^2 \, \chi \right)^{1/2} \right] \right\}^{1/2}
\end{eqnarray} 
where $\chi$ is the angle between the wave propagation
and magnetic field directions, $V_A \equiv B / {(4 \pi \, \rho)^{1/2}}$
is the scalar Alfv\'en speed, $c_s \equiv ( {\gamma p} / {\rho} )^{1/2}$
is the plasma sound speed, $c_{ms} \equiv ( V_A^2 + c_s^2 )^{1/2}$ is the
magnetosound speed, and $B$, $p$, $\rho$, and $\gamma$ are the magnetic
field strength, plasma pressure, mass density, and adiabatic index,
respectively.  Note: for the sake of straightforward discussion, we give here, 
and in equation (\ref{VF}) below, the non-relativistic versions of the slow 
and fast wave expressions.  The full relativistic versions (again in the 
fluid frame) are given in the Appendix and are used in our actual computations.  

Note that the propagation speed of the slow wave depends on whether
the magnetic forces internally dominate the plasma or vice-versa and
on the angle $\chi$ of propagation of the wave to the magnetic field.
In our magnetic model for BL Lac ($V_A^2 >> c_s^2$) the (non-relativistic)
slow wave speed is given by $V_S \approx (c_s V_A/c_{ms}) \cos \chi \sim
c_s \cos \chi$.  So, the slow wave is a simple sound wave ($V_S = c_s$)
{\em along} the magnetic field, but it propagates more slowly skew to the
field and has $V_S = 0$ normal to it.  Slow MHD shocks also can occur,
propagating at speeds faster than $V_S$.

Slow MHD waves propagating along the axis of the BL Lac jet through
the helical magnetic field will have a very slow speed in the frame
of the jet for two reasons.  First, the wave will be traveling skew to
the helical field, so its speed will be even less than the sound speed
(because $\cos \chi$ could be much less than unity); and, second, if
the field dominates, the sound speed could be quite low, so that $V_S
< c_s << V_A$.  Since slow MHD waves/shocks still compress the plasma
(but not the field), they should produce a visible enhancement in the
local synchrotron emission in the jet. However, that emission could be
almost stationary with respect to the moving jet and mainly reflect the
speed of the jet plasma beam when measured by the observer. Waves of
this type would conform to the idea that the moving components travel
with the beam, and that the Lorentz factor of the beam can be found from
the speed of the components.  This use of the slow magneto-acoustic mode
could provide a justification for saying that the pattern speed equals
the beam speed, in the proper circumstances.

\subsection{Fast MHD Waves and Shocks in the BL Lac Jet}

The upper branch of the magneto-acoustic modes, the fast MHD wave,
compresses not only the plasma but also the magnetic field itself.
Its propagation phase speed is given by the {\em positive} root of the
dispersion relation 
\begin{eqnarray} 
\label{VF} 
V_F & = & \pm \, \left\{
\frac{1}{2} \left[ c_{ms}^2 + \left(c_{ms}^4 - 4 \, c_s^2 \, V_A^2 \,
\cos^2 \, \chi \right)^{1/2} \right] \right\}^{1/2} 
\end{eqnarray} 
When the magnetic field dominates and the Alfv\'en speed exceeds $c_s$,
the speed of this wave is a very fast $V_F = V_A$ along the magnetic
field, and normal to the field an even faster $V_F = c_{ms}$.

In contrast to the slow mode, fast MHD waves have no problem propagating
along the jet axis at high speeds.  In fact, if the helical field is in
a fairly tight coil, $\cos^2 \chi$ will be small, and the fast wave speed
will approach the full magnetosound speed.  Furthermore, fast MHD shocks
with $V_{shock} > V_F$ are quite possible, most probably occurring when
a new, strong blast of plasma attempts to propagate up the magnetic coil.

Fast MHD waves, therefore, are likely to be responsible for the fastest
moving components in BL Lac.  In addition, because of their ability
to compress the fairly strong magnetic field, they fit well into the
model for synchrotron emission that has been applied to BL Lac by
\citet{HAA89}. The only difference is that the initial field here is
in an ordered helix, not in a tangled configuration.


\subsection{A Simple MHD Model for the Moving Components}
\label{simple model}


The considerable variation in the speeds of the components in
Figure~\ref{fig2} is commonly seen in superluminal sources \citepalias{Lis13}, and
the reasons usually given for this are that the components are due
to shocks of different strength, that $\theta$ is variable, or that
there are sub-beams of varying speed and direction.  Here we suggest
a slightly different model in which the speed of the waves or shocks
that form the components may vary, but largely because they are of two
different wave or shock types (slow and fast magneto-acoustic waves),
in addition to there being different speeds within the two classes. The
detailed distribution of speeds is presumed to be set by variability
at the RCS. Also, we assume that the speed of the beam and the viewing
angle $\theta$ are fixed, the latter at $6\deg$.

The model has two steps: (a) assume that the slowest speed in
Figure~\ref{fig2} results from a slow mode wave of negligible speed in
the jet frame and, from this, calculate the beam speed; (b) assume that
the fastest speed in Figure~\ref{fig2} results from a fast mode wave
traveling on the beam which, in turn, has the speed found in (a). This
gives the speed of the fastest fast wave.

The model has three important speeds:
$\rm \beta_{beam}^{gal}$, the speed of the beam in the galaxy frame,
and the speed of the wave in the
beam and galaxy frames,
$\rm \beta_{wave}^{beam}$ and
$\rm \beta_{wave}^{gal}$.
These are related by the relativistic velocity addition formula
\begin{eqnarray}
\label{beta_add}
{\rm\beta_{wave}^{gal}} & = & \frac{\rm\beta_{wave}^{beam}+\beta_{beam}^
{gal}} {\rm 1+\beta_{wave}^{beam} \beta_{beam}^{gal}}
\end{eqnarray}
In addition, $\rm \beta_{wave}^{gal}$ is related to the observed $\rm
\beta_{app}$ by the formula 
\begin{eqnarray}
\label{beta_app}
{\rm \beta_{app} = \frac{\beta_{wave}^{gal} \sin{\theta}} 
{1-\beta_{wave}^{gal}\cos{\theta}}}
\end{eqnarray}
and each speed has a corresponding Lorentz factor 
$\Gamma=(1-\beta^2)^{-1/2}$.

Now assume that the slowest speed in Figure~\ref{fig2} represents
the apparent speed of the slowest slow wave on the beam.  
From Figure~\ref{fig2} we choose the propagation of C5 from about 1999 to
2001 to represent the speed of the jet plus the slowest slow wave, or $\rm
\beta_{app,Swave}^{gal} = 2.1$. From Eqn~\ref{beta_app} with $\theta=6\deg$,
$\rm \beta_{Swave}^{gal} = 0.958$ and $\rm \Gamma_{Swave}^{gal}=3.5$.
We now assume the limiting case in which the slow wave speed
(Eqn~\ref{VS}) is small in the frame of the beam, or $\rm
\beta_{Swave}^{beam} \sim 0$.  This gives us the beam speed itself, $\rm
\beta_{beam}^{gal}\approx 0.958$ or $\rm \Gamma_{beam}^{gal}\approx 3.5$.
In this model, the slowest observed component indicates the actual speed
of the beam.

The magnetosound speed in the jet can then be found by assuming
that the fastest observed component speed is due to the fastest
fast magneto-acoustic wave, or $\rm \beta_{app,Fwave}^{gal}=10$.  Then,
from Eqn~\ref{beta_app}, $\rm \beta_{Fwave}^{gal} = 0.995$ and $\rm
\Gamma_{Fwave}^{gal}=10.1$.  Inverting Eqn~\ref{beta_add} and taking
$\rm \beta_{beam}^{gal}\approx 0.958$ gives $\rm \beta_{Fwave}^{beam}
\approx 0.793$ and $\rm \Gamma_{Fwave}^{beam}\approx 1.6$

In this model the beam is relativistic, $\rm \Gamma_{beam}^{gal}\approx
3.5$, the fast MHD wave is mildly relativistic in the beam frame,
$\rm \Gamma_{Fwave}^{beam}\approx 1.6$, and the slow MHD wave has
negligible speed relative to the plasma.  However, while this model
assumes that in the jet the internal magnetic forces dominate the
plasma forces ($V_A > c_s$), there is no information about the
configuration or orientation of the magnetic field relative to the
jet flow. This can be determined from the propagation of Alfv\'en
waves, but such waves are not considered in this paper.  They will,
however, be presented in Paper II.

The beam itself has $\rm \Gamma_{beam}^{gal}\approx 3.5$, a value lower
than has been found in some other studies; e.g. \citet{CAM13} who find
$\rm \Gamma_{beam}^{gal}\approx 5.4$.  However, our value is adequate to
suppress the flux density from the counter jet (see Figure~\ref{fig1}).
For $\rm \Gamma_{beam}^{gal}\approx 3.5$ and $\theta = 6\deg$ the Doppler
factor $\delta \approx 6$, and the jet to counter-jet flux-density ratio
is of order $\delta^{4-2\alpha} \sim 10^4$, taking the spectral index
of the jet as $\alpha = -0.55$ \citep{Hov14}.

Note that relativistic velocity addition is highly nonlinear. For
the fast wave, $\rm \Gamma_{Fwave}^{beam}\approx 1.6$ and
$\rm \Gamma_{beam}^{gal}\approx 3.5$; the combination gives $\rm
\Gamma_{Fwave}^{gal}\approx 10$. Here is a case where the Lorentz factor
for the pattern is 3 times that for the beam.  This is a general result
and does not depend on the strength of the magnetic field in the jet
model. For example, models for blazars with weak, tangled fields that
are compressed by propagating shocks also can have moderate beam Lorentz
factors with mildly relativistic shocks in the jet frame, and yet produce
very fast, highly relativistic component speeds in the observer's frame.


\section{Discussion}
\label{sec:discussion}

The results presented in this paper may have rather wide application,
because there are theoretical reasons for believing that all extragalactic
jets may have strong magnetic fields (relative to the kinetic pressure)
and a recollimation shock \citep{Mei13}. The objects in Table 1 are
close to the Earth so the linear resolution on them is good, and yet the
recollimation shock in BL Lac is at the limit of angular resolution at 15
GHz. Thus higher frequencies or space VLBI will be necessary to study many
more AGN, and to look for evidence of an RCS.  The characteristics to look
for in nearby objects are a stationary and bright feature (the candidate
RCS) $\lesssim 10^6\, r_g$ from the core, a transverse magnetic field in that
feature, and superluminal components that appear to emanate from the RCS.

We have included the class of each of the five objects in
Table~\ref{distancetable} in column 3.  Standard arguments (e.g. Chiaberge
et al. 2000) based on orientation unify BL Lacs with FR~I objects, and
flat-spectrum radio quasars (FSRQ) with FR~II objects.  Thus we consider
BL Lac, M\,87 and 3C\,120 as one group, and 3C\,273 and 3C\,390.3 as
another.  Column 8 of Table~\ref{distancetable} shows that the RCS in
the FSRQ/FR~II group is one to two orders of magnitude closer to the BH
than those in the BL Lac/FR~I group (component C80 in 3C\,120 is excluded
from this comparison; see \S\ref{sec:other_AGN}). Although the statistics
of this comparison are weak, the result is interesting and suggests that
the inner jets in the two groups of sources may differ in some way.
In addition, these groups are not magnetically homogeneous.  BL Lac has
longitudinal EVPA but 3C\,120 has transverse EVPA; \citet{Gom08} suggest
that this is caused by a helical magnetic field of low pitch angle.
M\,87 is unpolarized near the core at 15 GHz \citep{ZT02}. At HST-1 (the
RCS) the RM-corrected EVPA is longitudinal, but it is transverse between
the knots. 3C\,273 displays a transverse gradient in its RM \citep{ZT05,
Asa08, Hov12} that suggests it has a helical magnetic field whose pitch
angle is not small. Although the conventional picture says that this
should result in a longitudinal EVPA, in fact the EVPA is predominantly
transverse in 3C\,273.  For 3C\,390.3 any polarized flux is undetectable
i.e.\ weak \citep{THV01}, as it is in most radio galaxies \citep{LH05}.

In making the model we used in \S\ref{simple model} we assumed that the
speed of the beam is constant. This is unlikely to be true, but it is
convenient and provides a way to estimate the speeds of the beam and the
fast-mode wave. We can set a lower limit for the beam speed as follows.
BL Lac is strongly one-sided and the jet-to-counterjet flux-density
ratio is of order $10^3$ or more. The appropriate limit, assuming BL
Lac has symmetric opposite beams, is $\delta^{4-2\alpha} >1000$ where
$\delta$ is the Doppler factor and $\alpha = -0.55$, and so $\delta>3.9$
and $\rm\Gamma_{beam}>2$.  If we use the value $\theta = 6\deg$ then there
is an upper limit, found by noting that the MHD waves propagate downstream
on the beam, increasing the net speed seen by the observer. The fastest
observed speed is $\rm \beta_{app} = 10$, and with $\theta = 6\deg$ this
gives $\rm\Gamma_{beam} < 10$.  If we relax the condition $\theta=6\deg$,
then there is no upper limit within this geometric framework.

The observed pattern is faster than the beam, for a component generated by
a fast-mode MHD wave. For the particular model we have used, the fastest
observed superluminal component has $\rm \Gamma_{Fwave}^{gal}\approx 10$,
and $\rm\Gamma_{beam}^{gal}\approx 3.5$.  The pattern is approximately 3
times faster than the beam. This result has consequences for unification
theory, if it is confirmed.  The standard picture says that BL Lacs
and FR I galaxies form one class, with BL Lacs seen end-on and FR I's
seen closer to the equatorial plane. Analyses based on this idea have
difficulties, however, in that some properties, including luminosity
ratios and superluminal motion, over-predict the amount of radio flux
that should be seen in the compact components \citep{Chi00}.  This would
be alleviated if our model result is generally correct for BL Lacs, that
the Lorentz factor for the beam is smaller than the Lorentz factor for
the superluminal motion. On the other hand, such a general reduction in
Lorentz factor for the beams will also reduce the Doppler factors, and
that will exaggerate the difficulties that already exist in explaining
very high brightness temperatures \citep{Kov05, Sok13}.

\section{Conclusions}
\label{sec:conclusions}

The set of 110 images of BL Lac, observed with the VLBA at 15 GHz
between 1999.2 and 2013.0, provides a remarkable view of the jet of this
exceptional AGN. The jet has a strong quasi-stationary component (C7)
at $r\approx 0.26$ mas from the core.  Fast superluminal components
appear to emanate from C7, in a manner strikingly similar to the
ejection of fast shocks from the RCS that is seen in 2D RMHD numerical
simulations of jets that have a dominant, helical magnetic field. 
Furthermore, C7 is analogous to the component HST-1 in
M\,87, which has been called an RCS. We therefore identify C7 as an RCS
in the BL Lac jet. 

A simple model is employed, one that uses the observed slowest and fastest
moving components, with the assumption that they are manifestations
of slow and fast magneto-acoustic waves propagating downstream on the
relativistic beam. The model assumes that $\theta=6\deg$ and that the beam
speed is constant.  The result of fitting the model to the data is that
the Lorentz factor for the beam is $\rm \Gamma_{beam}^{gal}\approx 3.5$,
and that the fast magneto-acoustic wave has a speed relative to the beam
of $\rm \beta_{Fwave}^{beam}\approx 0.79$. This is mildly relativistic,
with $\rm \Gamma_{Fwave}^{beam}\approx 1.6$. These give the observed
pattern speed $\rm \beta_{app} = 10$.

In the model, the Lorentz factor of the observed pattern is approximately
3 times larger than the Lorentz factor of the beam. This difference is
in the correct sense to help alleviate difficulties in the BL Lac --
FR\,I unification; but it is in the wrong sense to reduce the difficulties
seen in explaining very high brightness temperatures.


\acknowledgments

We thank the anonymous referee whose comments substantially improved
the paper. We are grateful to Ken Kellermann and the rest of the MOJAVE
team for their comments and for their years of work in producing the data
base that makes this work possible.  TH was supported in part by a grant
from the Jenny and Antti Wihuri foundation and by the Academy of Finland
project number 267324.  YYK is partly supported by the Russian Foundation
for Basic Research (project 13-02-12103), Research Program OFN-17 of
the Division of Physics, Russian Academy of Sciences, and the Dynasty
Foundation.  ABP was supported by the ``Non-stationary processes in the
Universe'' Program of the Presidium of the Russian Academy of Sciences.
TGA acknowledges support by DFG project number Os 177/2-1. The VLBA is a
facility of the National Radio Astronomy Observatory, a facility of the
National Science Foundation that is operated under cooperative agreement
with Associated Universities, Inc.  The MOJAVE program is supported under
NASA-Fermi grant 11-Fermi11-0019.  Part of this research was carried out
at the Jet Propulsion Laboratory, California Institute of Technology,
under contract with the National Aeronautics and Space Administration.
This research has made use of NASA's Astrophysics Data System.



\appendix
\section{Relativistic Acoustic MHD Waves}

Here we present the relativistic expressions for the slow and fast
MHD wave counterparts to equations (\ref{VS}) and (\ref{VF}). They are
found in the same manner as the non-relativistic versions, but using
the relativistic MHD equations instead \citep[\S 9.5.2.1]{Mei12}: the
equations are linearized about zero-velocity flow with all quantities
having small perturbations proportional to $\exp[i({\bm k} \cdot {\bm x}
- \omega t)]$; keeping only acoustic perturbations (${\bm k} \times {\bm
\delta V} = 0$) produces a quadratic equation in $V^2$ with two roots.
Using the same form as the non-relativistic versions in the text, the
lower root gives the slow wave relativistic speed
\begin{eqnarray}
\label{rel_VS}
\nonumber
\beta_S & = & \pm \left\{ \frac{1}{2} \left[ \beta_{ms}^2 \left\{1 + \beta_c^2 \cos^2 \chi\right\} \right. \right.
\\
& & \left. \left. - \left( \beta_{ms}^4 \left\{1 + \beta_c^2 \cos^2 \chi\right\}^2 \, - \, 
4 \, \beta_s^2 \beta_A^2 \, \cos^2 \chi \right)^{1/2} \right] \right\}^{1/2}
\end{eqnarray}
and the upper root gives the fast wave relativistic speed
\begin{eqnarray}
\label{rel_VF}
\nonumber
\beta_F & = & \pm \left\{ \frac{1}{2} \left[ \beta_{ms}^2 \left\{1 + 
\beta_c^2 \cos^2 \chi\right\} \right. \right.
\\
& & \left. \left. + \left( \beta_{ms}^4 \left\{1 + \beta_c^2 \cos^2 
\chi\right\}^2 \, - \, 
4 \, \beta_s^2 \beta_A^2 \, \cos^2 \chi \right)^{1/2} \right] \right\}^{1/2}
\end{eqnarray}
\citep[equations 9.192--9.196]{Mei12}, where $\chi$ is the angle between
the propagation direction ${\bm k}$ and the magnetic field direction.
Note that $\beta_S$ is the relativistic counterpart to the slow wave
speed $V_S$ in \S4, while $\beta_s$ is the counterpart to the plasma
sound speed $c_s$.  These expressions reduce to the non-relativistic
equations (\ref{VS}) and (\ref{VF}) when terms of order higher than
$\beta^2$ in the root (and higher than $\beta^4$ in the discriminant)
are discarded.  The relativistic cusp and magnetosound speeds in the above are
\begin{eqnarray}
\label{rel_Vcusp}
\beta_c & = & \beta_s \, \beta_A / \beta_{ms}
\\
\label{rel_Vms}
\beta_{ms} & = & \left[ \beta_A^2 \, + \, \beta_s^2 \, \left(1 - 
\beta_A^2 \right) \right]^{1/2}
\end{eqnarray}
and the relativistic adiabatic sound and Alfv\'en speeds are 
\begin{eqnarray}
\label{rel_Vs_VA}
\beta_s & = & \left( \frac{\gamma \, p}{\rho \, c^2 + \varepsilon + p} \right)^{1/2}
\\
\beta_A & = & \frac{B}{ \left[ 4 \pi \left( \rho \, c^2 + \varepsilon + 
p + B^2 / 4 \pi \right) \right]^{1/2}}
\end{eqnarray}
where $\varepsilon$ is the internal energy of the plasma; other quantities
are defined in \S \ref{move_comp_mhd}.

There are a few points of note.  While the Alfv\'en speed $\beta_A
\rightarrow 1$ as the magnetic pressure $B^2/8\pi \rightarrow \infty$,
the same is not true for the adiabatic sound speed of a relativistic
plasma ($\gamma = 4/3$, $\varepsilon = 3 p$, $\rho \, c^2 << p$):
as $p \rightarrow \infty$, $\beta_s \rightarrow (1/3)^{1/2} \approx
0.577$ ($\Gamma_s \sim 1.22$).  
Sound waves in relativistic jets, therefore, may be limited to modest
Lorentz factors. Note also the limitation in equation (\ref{rel_Vms})
on the sound speed contribution to the magnetosound speed when the
magnetic field is strong and the Alfv\'en speed relativistic. In that
case $\beta_{ms} \approx \beta_A$ and $\beta_c \approx \beta_s$ are
good approximations.  In the limit $\beta_A \rightarrow 1$, we always
have $\beta_{ms} \rightarrow 1$, no matter what the value of $\beta_s$.

Finally, while expressions (\ref{rel_VS}) and (\ref{rel_VF}) may look
like they could exceed unity (and violate relativistic principles),
they do not. The extrema of these quantities occur when $\beta_{ms} =
\beta_A = 1$, for any value of $\chi$, and they are given by $\beta_S =
\pm \beta_s \cos \chi$ and $\beta_F = \pm 1$.


\begin{thebibliography}{}

\bibitem[Agudo et al.(2012)]{Agu12}
Agudo, I., G\'omez, J.-L., Casadio, C., Cawthorne, T.V. \& Roca-Sogorb, M.
2012, \apj, 752, 92

\bibitem[Arshakian et al.(2010)]{Ars10} 
Arshakian, T.~G., Le\'on-Tavares, J., Lobanov, A.~P., et al.
2010, \mnras, 401, 1231

\bibitem[Asada et al.(2002)]{Asa02} Asada, K., Inoue, M., Uchida, Y. et
al. 2002, \pasj, 54, L39

\bibitem[Asada et al.(2008)]{Asa08}
Asada, K., Inoue, M., Kameno, S. \& Nagai, H. 2008, \apj, 675, 79

\bibitem[Bach et~al.(2006)]{Bac06}
Bach, U., Villata, M., Raiteri, C.~M., et~al. 2006, \aap, 456, 105

\bibitem[Blandford(1993)]{Bla93} Blandford, R. 1993, in
Astrophysical Jets, Astrophysics and Space Science Library (Cambridge:
Cambridge Univ. Press), Vol. 103, 15

\bibitem[{{Blandford} \& {Payne}(1982)}]{BP82}
{Blandford}, R.~D., \& {Payne}, D.~G. 1982, \mnras, 199, 883

\bibitem[{{Bromberg} \& {Levinson}(2009)}]{BL09}
{Bromberg}, O., \& {Levinson}, A. 2009, \apj, 699, 1274

\bibitem[{{Caproni} {et~al.}(2013){Caproni}, {Abraham}, \& {Monteiro}}]{CAM13}
{Caproni}, A., {Abraham}, Z., \& {Monteiro}, H. 2013, \mnras, 428, 280

\bibitem[Chang et al.(2010)]{Cha10} Chang, C.~S., Ros,
E., Kovalev, Y.~Y., \& Lister, M.~L.\ 2010, \aap, 515, A38

\bibitem[{{Cheung} {et~al.}(2007){Cheung}, {Harris}, \& {Stawarz}}]{Che07}
{Cheung}, C.~C., {Harris}, D.~E., \& {Stawarz}, {\L}. 2007, \apjl, 663, L65

\bibitem[{{Chiaberge} {et~al.}(2000){Chiaberge}, {Celotti}, {Capetti}, \&
{Ghisellini}}]{Chi00}
{Chiaberge}, M., {Celotti}, A., {Capetti}, A., \& {Ghisellini}, G. 2000, 
\aap, 358, 104

\bibitem[{{Clarke} {et~al.}(1986){Clarke}, {Norman}, \& {Burns}}]{CNB86}
{Clarke}, D.~A., {Norman}, M.~L., \& {Burns}, J.~O. 1986, \apjl, 311, L63

\bibitem[Cohen et al.(2007)]{Coh07}
Cohen, M.~H., Lister, M.~L., Homan, D.~C., et al. 2007, \apj, 658, 232

\bibitem[{{Falomo} {et~al.}(2002){Falomo}, {Kotilainen}, \& {Treves}}]{FKT02}
{Falomo}, R., {Kotilainen}, J.~K., \& {Treves}, A. 2002, \apjl, 569, L35

\bibitem[Gabuzda et~al.(2004)]{GMC04}
Gabuzda, D.~C., Murray, E., \& Cronin, P. 2004, \mnras, 351, L89

\bibitem[Gabuzda et~al.(2000)]{GPC00}
Gabuzda, D.~C., Pushkarev, A.~B. \& Cawthorne, T.~V. 2000, 
\mnras, 319, 1109

\bibitem[{{Gebhardt} {et~al.}(2011){Gebhardt}, {Adams}, {Richstone}, {Lauer},
  {Faber}, {G{\"u}ltekin}, {Murphy}, \& {Tremaine}}]{Geb11}
{Gebhardt}, K., {Adams}, J., {Richstone}, D., {et~al.}, 2011, \apj, 729, 119

\bibitem[G\'omez et~al.(2001)]{Gom01}
G\'omez, J.-L., Marscher, A.~P., Alberdi, A., Jorstad, S. \& Agudo,
I. 2001, \apj, 561, L161

\bibitem[G\'omez et~al.(2008)]{Gom08}
G\'omez, J.-L., Marscher, A.~P., Jorstad, S.~G, Agudo, I
\& Roca-Sogorb, M. 2008, \apj, 681, L72

\bibitem[Grier et al.(2012)]{Gri12}
Grier, C.~J., Peterson, B.~M., Pogge, R.~W., et al., 2012, \apj, 755, 60

\bibitem[{{Homan} {et~al.}(2009){Homan}, {Kadler}, {Kellermann}, {Kovalev},
  {Lister}, {Ros}, {Savolainen}, \& {Zensus}}]{Hom09}
{Homan}, D.~C., {Kadler}, M., {Kellermann}, K.~I., 
{et~al.} 2009, \apj, 706, 1253

\bibitem[{{Hovatta} {et~al.}(2009){Hovatta}, {Valtaoja}, {Tornikoski},
\& {L{\"a}hteenm{\"a}ki}}]{Hov09} {Hovatta}, T., {Valtaoja}, E.,
{Tornikoski}, M., \& {L{\"a}hteenm{\"a}ki}, A.  2009, \aap, 494, 527

\bibitem[Hovatta et~al.(2012)]{Hov12}
Hovatta, T., Lister, M.~L., Aller, M.~F., et~al. 2012, \aj, 144, 105

\bibitem[Hovatta et~al.(2014)]{Hov14}
Hovatta, T. et~al. 2014, \aj, in press; arXiv:1404.0014

\bibitem[{{Hughes} {et~al.}(1989){Hughes}, {Aller}, \& {Aller}}]{HAA89}
{Hughes}, P.~A., {Aller}, H.~D., \& {Aller}, M.~F. 1989, \apj, 341, 68

\bibitem[Jorstad et~al.(2005)]{J05}
Jorstad, S.~G., Marscher, A.~P., Lister, M.~L., et~al. 2005, 
\aj, 130, 1418
\citepalias{J05}

\bibitem[{{Komissarov}(1999)}]{Kom99}
{Komissarov}, S.~S. 1999, \mnras, 308, 1069

\bibitem[Kovalev et al.(2005)]{Kov05} Kovalev, Y.~Y.,
Kellermann, K.~I., Lister, M.~L., et al. 2005, \aj, 130, 2473 

\bibitem[Kovalev et al.(2008)]{Kov08}
Kovalev, Y.~Y., Lobanov, A.~P., Pushkarev, A.~B., \& Zensus, J.~A.\
2008, \aap, 483, 759

\bibitem[{{Krause} \& {Camenzind}(2001)}]{KC01}
{Krause}, M., \& {Camenzind}, M. 2001, \aap, 380, 789

\bibitem[Le\'on-Tavares et al.(2010)]{Leo10}
Le\'on-Tavares, J., Lobanov, A.~P., Chavushyan, V.~H., 
 et~al. 2010, \apj, 715, 355

\bibitem[Li et~al.(1992)]{LCB92}
Li, Z.-Y., Chiueh, T., \& Begelman, M.~C. 1992, \apj, 394, 459

\bibitem[{{Lind} {et~al.}(1989){Lind}, {Payne}, {Meier}, \&
{Blandford}}]{Lin89}
{Lind}, K.~R., {Payne}, D.~G., {Meier}, D.~L., \& {Blandford}, R.~D. 1989,
\apj, 344, 89

\bibitem[Lister \& Homan(2005)]{LH05}
Lister, M.~L., \& Homan, D.~C. 2005, \aj, 130, 1389

\bibitem[Lister \& Marscher(1997)]{LM97}
Lister, M.~L., \& Marscher, A.~P. 1997, \apj, 476, 572

\bibitem[Lister et al.(2009)]{Lis09} Lister, M.~L.,
Cohen, M.~H., Homan, D.~C., et al.\ 2009, \aj, 138, 1874

\bibitem[Lister et al.(2013)]{Lis13}
Lister, M.~L., Aller,M.~F., Aller, H.~D., et al. 2013, \aj, 146, 120 
\citepalias{Lis13}

\bibitem[Lyutikov et~al.(2005)]{LPG05}
Lyutikov, M., Pariev, V.~I. \& Gabuzda, D.~C. 2005, \mnras, 360, 869

\bibitem[Marscher et al.(2008)]{Mar08}
Marscher, A.~P., Jorstad, S.~G., D'Arcangelo, F.~D., et al. 2008, \nat, 452, 966

\bibitem[{{Meier}(2012)}]{Mei12}
{Meier}, D.~L. 2012, {``Black Hole Astrophysics: The Engine Paradigm"},
Springer, 2012

\bibitem[Meier(2013)]{Mei13}
Meier, D. 2013 in ``The Innermost Regions of Relativistic Jets and Their
Magnetic Fields", EPJ Web of Conferences, 61, 01001

\bibitem[{{Mo{\'o}r} {et~al.}(2011){Mo{\'o}r}, {Frey}, {Lambert}, {Titov}, \&
{Bakos}}]{Moo11}
{Mo{\'o}r}, A., {Frey}, S., {Lambert}, S.~B., {Titov}, O.~A., \& {Bakos}, J.
2011, \aj, 141, 178

\bibitem[{{Mutel} \& {Denn}(2005)}]{MD05}
{Mutel}, R.~L., \& {Denn}, G.~R. 2005, \apj, 623, 79 (MD05)

\bibitem[O'Sullivan \& Gabuzda(2009a)]{OG09a}
O'Sullivan, S.~P., \& Gabuzda, D.~C. 2009a, \mnras, 393, 429 

\bibitem[O'Sullivan \& Gabuzda(2009b)]{OG09b}
---. 2009b, \mnras, 400, 26

\bibitem[Paltani \& T\"urler(2005)]{PT05}
Paltani S. \& T\"urler, M. 2005, \aap, 435, 811

\bibitem[{{Polko} {et~al.}(2010){Polko}, {Meier}, \& {Markoff}}]{PMM10}
{Polko}, P., {Meier}, D.~L., \& {Markoff}, S. 2010, \apj, 723, 1343

\bibitem[{{Polko} {et~al.}(2013a){Polko}, {Meier}, \& {Markoff}}]{PMM13a}
---. 2013, \mnras, 428, 587

\bibitem[Polko et~al.(2013b)]{PMM13b}
---. 2014, \mnras, 438, 959

\bibitem[Pushkarev et al.(2009)]{Pus09}
Pushkarev, A.~B., Lister, M.~L., Kovalev, Y.~Y., \& Savolainen, T. 2009,
\aap, 507, L33


\bibitem[Pushkarev et al.(2012)]{Pus12} Pushkarev, A.~B., Hovatta, T.,
Kovalev, Y.~Y., et al.\ 2012, \aap, 545, A113

\bibitem[{{Savolainen} {et~al.}(2006){Savolainen}, {Wiik}, {Valtaoja}, \&
  {Tornikoski}}]{Sav06}
{Savolainen}, T., {Wiik}, K., {Valtaoja}, E., \& {Tornikoski}, M. 2006, \aap,
  446, 71

\bibitem[{{Sokolovsky} {et~al.}(2011){Sokolovsky}, {Kovalev}, {Pushkarev},
\& {Lobanov}}]{Sok11} {Sokolovsky}, K.~V., {Kovalev}, Y.~Y., {Pushkarev},
A.~B., \& {Lobanov}, A.~P. 2011, \aap, 532, A38

\bibitem[Sokolovsky(2013)]{Sok13}
Sokolovsky, K.~V. 2013, Proceedings of Science, 11th EVN Symposium, 109

\bibitem[{{Stawarz} {et~al.}(2006){Stawarz}, {Aharonian}, {Kataoka},
  {Ostrowski}, {Siemiginowska}, \& {Sikora}}]{Sta06}
{Stawarz}, {\L}., {Aharonian}, F., {Kataoka}, J.,
{et~al.} 2006, \mnras, 370, 981

\bibitem[Stirling et~al.(2003)]{Sti03}
Stirling, A.~M., Cawthorne, T.V., Stevens, J.~A., et~al. 2003, 
\mnras, 341, 405 (S03)

\bibitem[{{Tateyama}(2009)}]{Tat09}
{Tateyama}, C.~E. 2009, \apj, 705, 877

\bibitem[Taylor et~al.(2001)]{THV01}
Taylor, G.~B., Hough, D.~H. \& Venturi, T. 2001 \apj, 559, 703

\bibitem[Taylor \& Zavala(2010)]{TZ10}
Taylor, A.~R. \& Zavala, R. 2010, \apjl, 722, L183

\bibitem[{{Vlahakis} \& {K{\"o}nigl}(2003)}]{VK03}
{Vlahakis}, N., \& {K{\"o}nigl}, A. 2003, \apj, 596, 1080

\bibitem[{{Vlahakis} {et~al.}(2000){Vlahakis}, {Tsinganos}, {Sauty}, \&
{Trussoni}}]{Vla00}
{Vlahakis}, N., {Tsinganos}, K., {Sauty}, C., \& {Trussoni}, E. 2000,
\mnras, 318, 417

\bibitem[Wandel et al.(1999)]{WPM99}
Wandel A., Peterson B.~M. \& Malkan M.~A. 1999, \apj, 526, 579

\bibitem[Woo \& Urry(2002)]{WU02}
Woo, J.-H. \& Urry, C.~M. 2002, \apj, 579, 530

\bibitem[Zamaninasab et~al.(2013)]{Zam13}
Zamaninasab, M., Savolainen, T., Clausen-Brown, E. et~al. 2013,
\mnras, 436, 3341

\bibitem[Zavala \& Taylor(2002)]{ZT02}
Zavala, R.~T. \& Taylor, G.~B. 2002, \apj, 566, L9

\bibitem[Zavala \& Taylor(2005)]{ZT05} 
---. 2005, \apj, 626, L73

\end{thebibliography}
\end{document}